# Revisiting the secondary climate attributes for transportation infrastructure management: A Redux and Update for 2020

Tao Liao[1], Paul Kepley[2], Indraneel Kumar[3], Samuel Labi[4]

**ABSTRACT**
Environmental conditions in various regions can have a severely negative impact on the longevity and durability of the civil engineering infrastructures. In 2018, a published paper used 1971 to 2010 NOAA data from the contiguous United States to examine the temporal changes in secondary climate attributes (freeze-thaw cycles and freeze index) using the climate normals from two time windows, 1971-2000 and 1981-2010. That paper investigated whether there have been statistically significant changes in climate attribute levels across the two time windows, and used GIS-based interpolation methods to develop isarithmic maps of the climate attributes to facilitate their interpretation and application. This paper updates that study. In this study, we use NOAA climatic data from 1991 to 2020 to construct continuous surface maps showing the values of the primary and secondary climate attributes at the 48 continental states. The new maps provide an updated picture of the freeze index values and freeze-thaw cycles for the most recent climate condition. These new values provide a better picture of the freezing season characteristics of the United States, and will provide information necessary for better winter maintenance procedures and infrastructure design to accommodate regional climate differences.

## 1. INTRODUCTION
The influence of climate on the physical condition of transportation infrastructure is well documented in the literature. For road pavement infrastructure, for example, the share of non-load, mostly climatic, factors of pavement deterioration and subsequent pavement repair costs typically ranges from 20%-60% compared to those from traffic load factors, depending on the type of pavement (Li, Sinha and McCarthy 2007, Fwa and Sinha 1987). The impacts of climate on infrastructure have been determined to occur through a variety of mechanisms including direct degradation of the infrastructure material and disruption of the stability or mechanical properties of the soil in which the structure is founded. Direct degradation through climatic influences may be classified by the process of deterioration (physical or chemical) or the rate of deterioration. The objective of this study is to use current data from more than 6,000 NOAA stations to provide updated climate maps for the continental states of the U.S. based on freeze-thaw cycles, freeze index, precipitation, and temperature.

## 2. MEASURES OF CLIMATIC SEVERITY

### 2.1 Primary Measures

*2.1(a) Temperature*

1. Independent Researcher, St. Louis, MO 63106, email: liaotao11@gmail.com, Corresponding Author; 2. Independent Researcher, Chicago, IL 60601; 3. Principal Regional Planner ; Purdue Center for Regional Development, The Schowe House, 1341 Northwestern Avenue , West Lafayette IN, 47907, Ph: (765)494-9485; Email: ikumar@purdue.edu; 4. Professor, Lyles School of Civil Engineering, Purdue University, 550 Stadium Mall Drive, West Lafayette IN, 47907, Ph: (765) 494-5926, Email: labi@purdue.edu

Physical impact as a result of climatic conditions include the effect of extreme heat or cold on bitumen: the rheological properties of bitumen render the stability of asphaltic concrete pavements susceptible to accelerated failure at temperature extremes. Under high temperatures, bitumen becomes increasingly viscous and undergoes plastic deformation (Bhattacharjee, Mallick and Daniel 2008). For this reason, rutting failures are generally more pronounced in the warmer regions (Hadley 1994). On the other extreme, low temperatures cause asphaltic cement to become brittle and vulnerable to cracking under traffic loading (Allen, Berg and Bigl 1991). For this reason, asphalt pavement cracking is generally more common at the colder regions. Compared to asphaltic concrete pavements, Portland cement concrete pavements seem to generally suffer relatively less damage due to temperature extremes, but appear to be more vulnerable to temperature changes. Expansion and contraction forces in concrete due to temperature variations cause stresses in concrete pavement that ultimately lead to failure (Kerr and Shade 1984).For both flexible and rigid pavements, the underlying subgrade is vulnerable to temperature variations. Thawing in the spring season causes subgrades to lose strength, and this effect is especially pronounced when ice lenses are present in the subgrade (Yoder and Witczak 1975). For this reason, pavement deterioration is most severe in the spring season (Allen, Berg and Bigl 1991) particularly under heavy truck loads.

*2.1(b) Precipitation*

Another aspect of the weather that accelerates pavement deterioration is precipitation: subgrade heave, resulting from expansion of clayey soils in response to increase in moisture content (arising, in part, from recharge due to precipitation), is a major cause of longitudinal cracking and other crack types on pavement surfaces. In addition, all else being equal, pavements in regions with higher precipitation suffer more deterioration because more water enters the underlying pavement layers through any surface cracks and unsealed or deteriorated joints. This can lead to pothole formation, deterioration of existing patches, ejection of a water-fines slurry (pumping) and subsequent void formation under the pavement, and loss of base or subgrade support (Yoder and Witczak 1975).

*2.1(c) Depth of Frost Penetration*

Figure 1 shows the depth of frost penetration, a major determinant in pavement failure. The freeze index (a measure of severity of frost in a region, in degree-days) varies considerably across the US. These trends suggest that there are significant variations in weather patterns across the US to warrant division of the contiguous part of the country into zones on the basis of climate conditions.

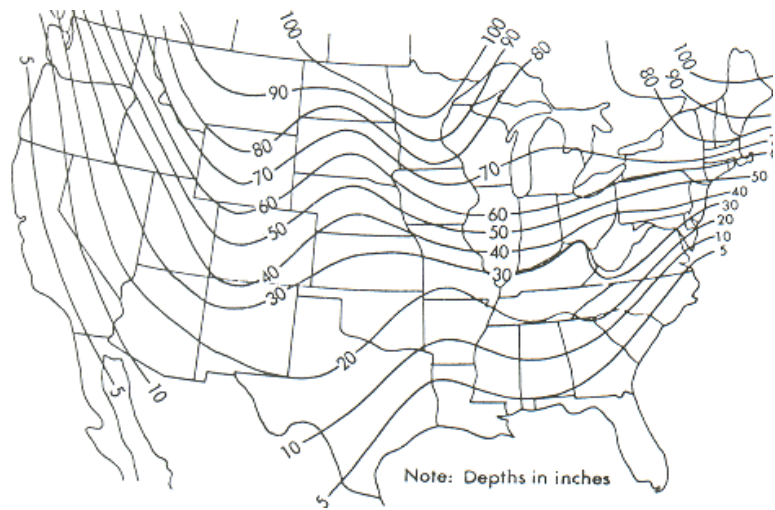

*Figure 1: Variation of the depth of frost penetration in inches (Yoder and Witczak 1975)*

**2.2 Secondary Measures**

This section presents a review of past research on the development of secondary indices of climatic severity that are derived from the primary measures discussed in the previous section of this paper. The secondary measures were developed for purposes of infrastructure damage studies. There are three ways in which the effect of climate could be considered in infrastructure damage studies: using a dummy variable representing climatic region; using disaggregate climate parameters such as freeze index, or by using a single aggregate index that embodies the effect of all climatic factors.

*2.2 (a) Freeze Index (FI)*
The freeze index has been shown to have a detrimental effect on joint deterioration and traverse joint faulting in rigid pavement, as well as causing medium to severe joint spalling in both plain concrete pavements and reinforced concrete pavements (Huang 2004). It has also been shown that the freeze index can be used to determine the frost penetration of a particular region, which can cause possible loss of soil compaction and non-uniform heave in pavements and other structures (Yoder and Witczak 1975, U.S. Army Corps of Engineers 1962). Figure 2 shows the distribution of freeze index in the continental US, developed by the Army Corps of Engineers in 1959.

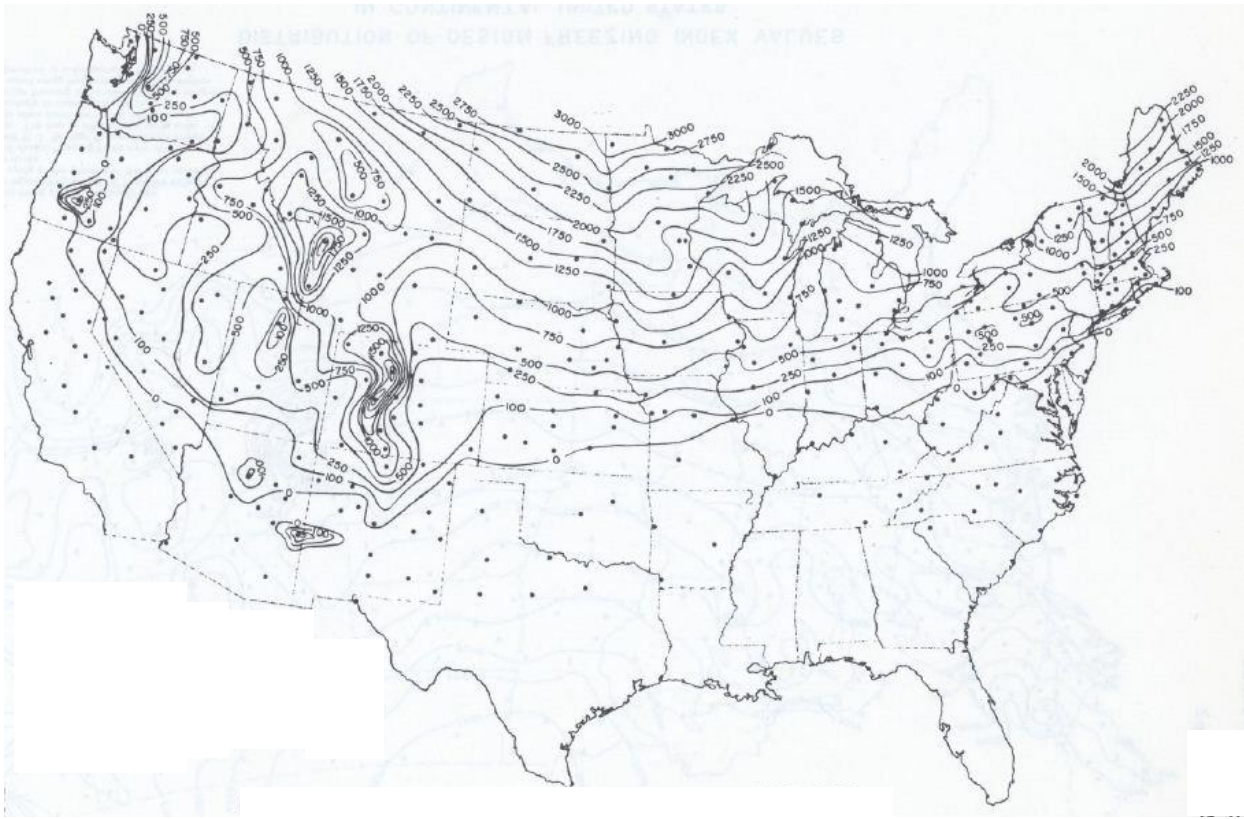

*Figure 2: Distribution of freeze index of continental U.S. - 1959 (U.S. Army Corps of Engineers 1962)*

*2.2(b) Freeze-Thaw Cycles (FTC)*
Freezing conditions during the winter season has been found to cause considerable problems in concrete and pavement infrastructures, because they alter the geotechnical properties of soil due to the physical effects that result from such conditions. It has been shown that freeze-thaw cycles cause increase in hydraulic permeability of soil due to crack development and structural degradation of fine grained soils such as clay (Qi., Vermeer and Cheng 2006). In addition, when exposed to the combined effects of loading, freeze-thaw cycles, and chloride from the de-icing salt, the deterioration in concrete was accelerated due to loss in weight and loss in dynamic modulus of elasticity (Mu, et al. 2002). A study carried out on fiber-reinforced polymer has also shown that when exposed to freeze-thaw cycles, the mode of failure was changed from adhesive to combined adhesive/cohesive (Lopez-Anido, Michael and Sandford 2004). Figure 3 shows the distribution of freeze-thaw cycles in the continental US, developed by Hershfield (Hershfield 1974) in 1974.

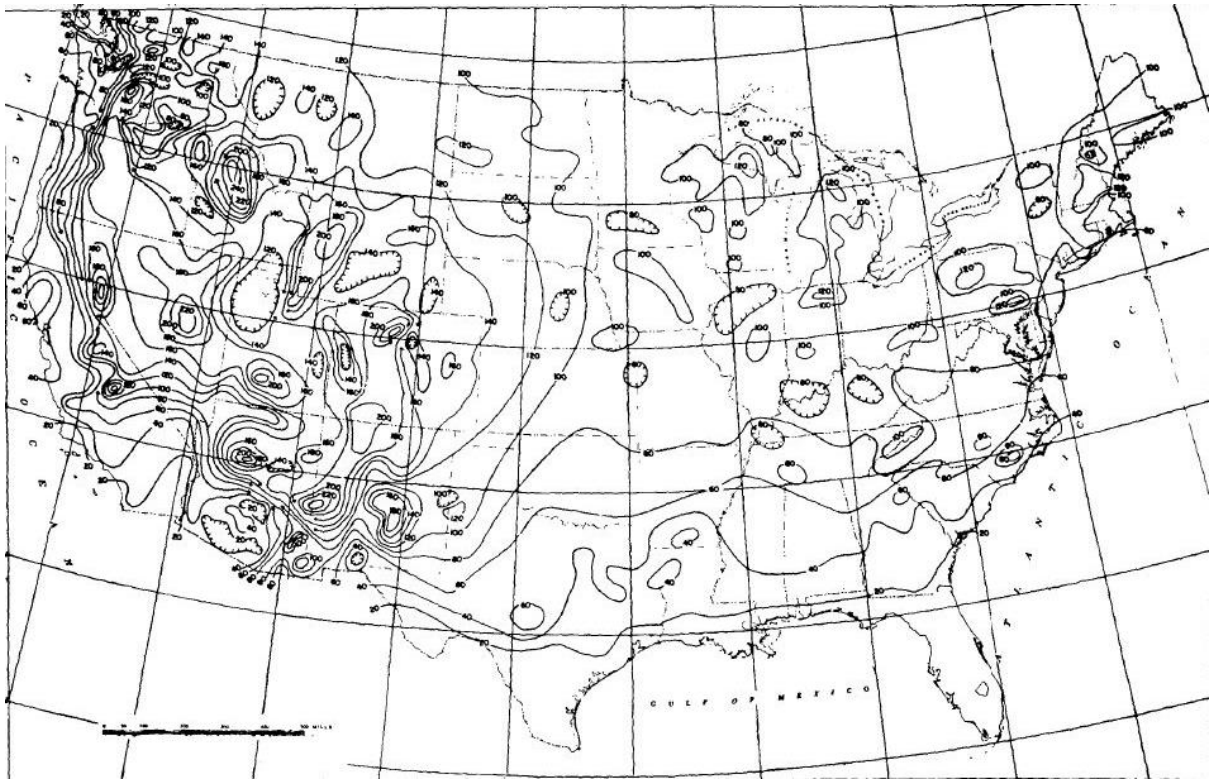

*Figure 3: Freeze-thaw cycle of the continental US – 1974 (Hershfield 1974)*

**DATA AND METHODOLOGY**

This follow up study uses the recently released 1991-2020 climate normal, as well as 1971-2000 and 1981-2010 climate normal from the previous study (Arguez, et al. 2012). The 1971-2000 and 1981-2010 climate normal contain approximately 2,200 weather stations each, while the 1991-2020 climate normal contains over 6,000 weather stations, greatly enhancing the spatial resolution, particularly in the western states. The previous study used roughly one weather station per county, the current study uses nearly all available weather stations that have the complete set of data. The continental US has 3,108 counties. As with the past study, the primary attributes (precipitation and temperature) and secondary attributes (freeze-thaw cycle and freeze index) were obtained or derived from the climate normal, and the data had gone through quality assurance procedure prior to publication by NOAA (Durre, et al. 2010).

For this study, the freeze index is defined as the magnitude (degrees) and duration (days) of "coldness" (below freezing air temperature) experienced in a given season. It can be calculated as the difference between the highest point and the lowest point on a cumulative degree-day curve, and a degree-day is the difference between the average daily air temperature and 32°F. A freeze-thaw cycle is defined as having occurred if the daily temperature was a maximum temperature that was more than 32°F and a minimum temperature that was less than 32°F. (Liao, et al. 2018)

This paper also uses the same Inverse Distance Weighting (IDW) interpolation method as in the previous (2018) study. The IDW interpolation assumes that the measured values closest to the prediction location carry more weight compared to those located farther away; thus, it assumes that the influence of each measured value diminishes with distance (Lam 1983).

The IDW method is calculated as follows:

$$p_i = \sum_{j=1}^{N} p_j/d^n{}_{ij} \div \sum_{j=1}^{N} 1/d^n{}_{ij} \qquad (1)$$

Where $p_i$ is the predicted value at point $i$; $p_j$ is the value at sampled point $j$, weather station data; $d_{ij}$ is the distance from point $i$ to point $j$; $N$ is the number of existing sampled values; and $n$ is inverse-distance weighting power. The value $n$ controls the sphere of influence for each of the existing sampled values. In this study, $n$ is set to equal 2, as this value was used in a previous study to interpolate rainfall data (NOAA 1972). Statistical significance between 1981-2010 and 1990-2020 time windows was determined using one-factor Analysis of Variance (ANOVA).

Similar to the previous study, NOAA climate regions are used in this (Arguez, et al. 2012) study to further examine the climate trends. NOAA climate regions have nine regions: Northwest (Idaho, Oregon, Washington); West (California, Nevada), Northern Rockies and Plains (Montana, Nebraska, North Dakota, South Dakota, Wyoming); Southwest (Arizona, Colorado, New Mexico, Utah); Upper Midwest (Iowa, Michigan, Minnesota, Wisconsin); Ohio Valley (Illinois, Indiana, Kentucky, Missouri, Ohio, Tennessee, West Virginia); South (Arkansas, Kansas, Louisiana, Mississippi, Oklahoma, Texas); Southeast (Alabama, Florida, Georgia, North Carolina, South Carolina, Virginia); and Northeast (Connecticut, Delaware, Maine, Maryland, Massachusetts, New Hampshire, New Jersey, New York, Pennsylvania, Rhode Island, Vermont). These regions were established by NOAA on the basis of their consistent climatic characteristics. (Karl and Koss 1984)

**RESULTS**

Figures 5 presents the isarithmic surface maps of the climate attributes of freeze index, freeze-thaw cycle, temperature, and precipitation for 1991-2020 normal. Isarithmic mapping was used because the data collected in the weather stations are continuous in nature and these maps can be used to develop isopleths or contour lines. Figure 6 presents the climate attribute in 1991-2020 time window. To better compare the climate trends in the three time windows (1971-2000, 1981-2010, 1999-2020), the 1991-2020 weather stations were filtered to find the geographically nearest-neighbor to match the past data. Figure 7 shows the climate comparison throughout all three time windows using only the nearest-neighbor weather stations in 1991-2020 climate normal.

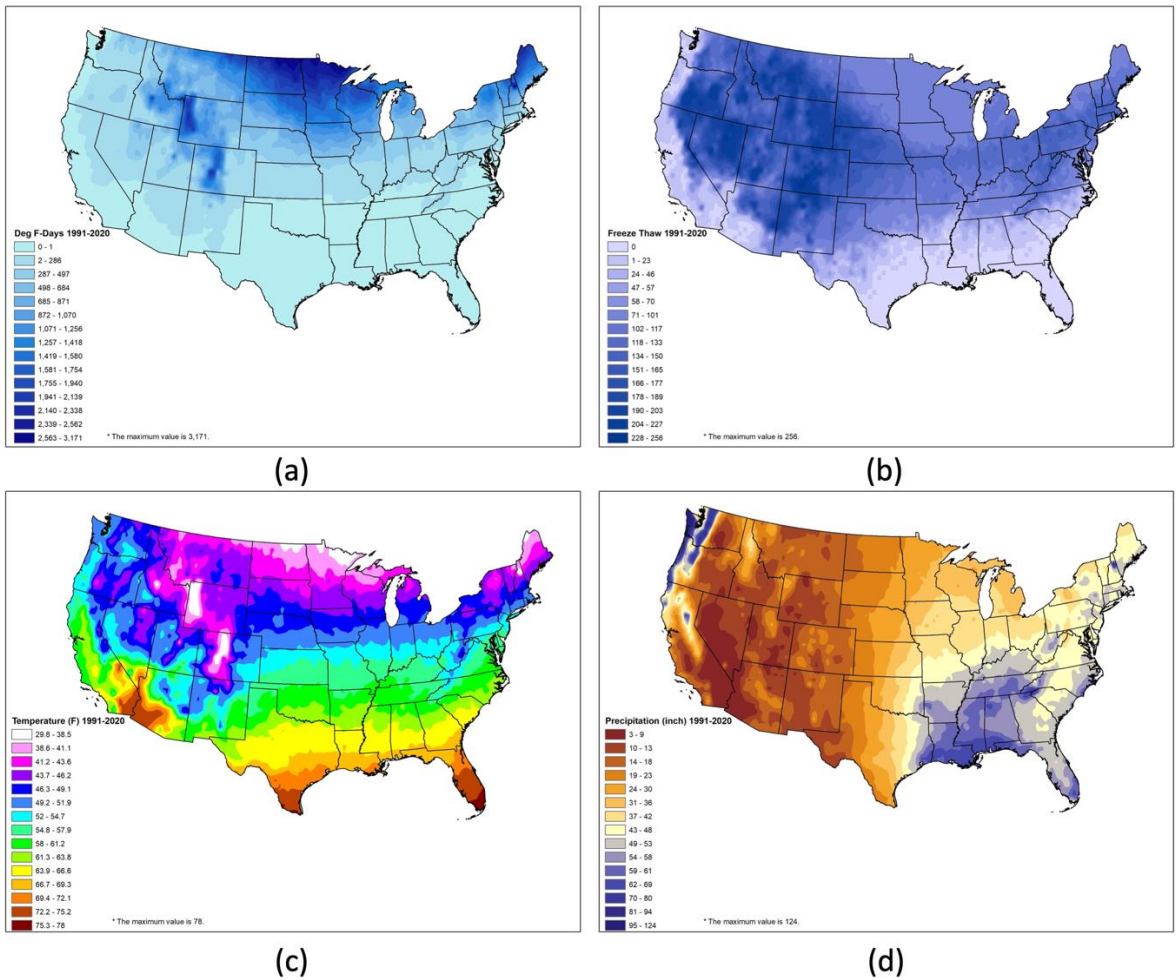

*Figure 4: (a) average annual freeze index, 1991-2020; (b) average annual number of freeze-thaw cycles, 1991-2020; (c) average annual temperature, 1991-2020; (d) average annual precipitation, 1991-2020*

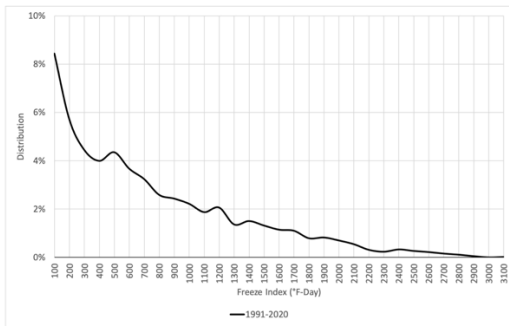 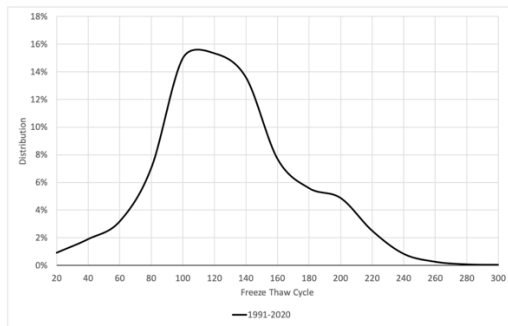
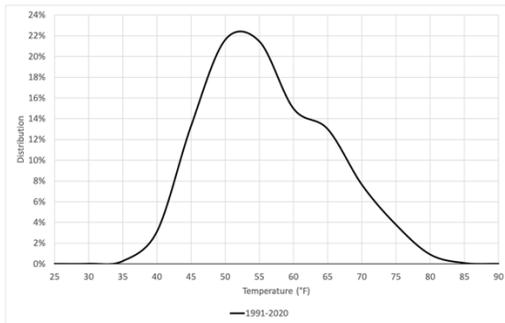 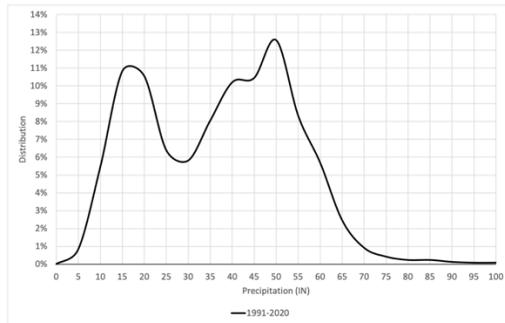

*Figure 5: (a) Freeze index trend, 1991-2020; (b) Freeze-thaw cycle trend, 1991-2020; (c) Temperature trend, 1991-2020; (d) Precipitation trend, 1991-2020*

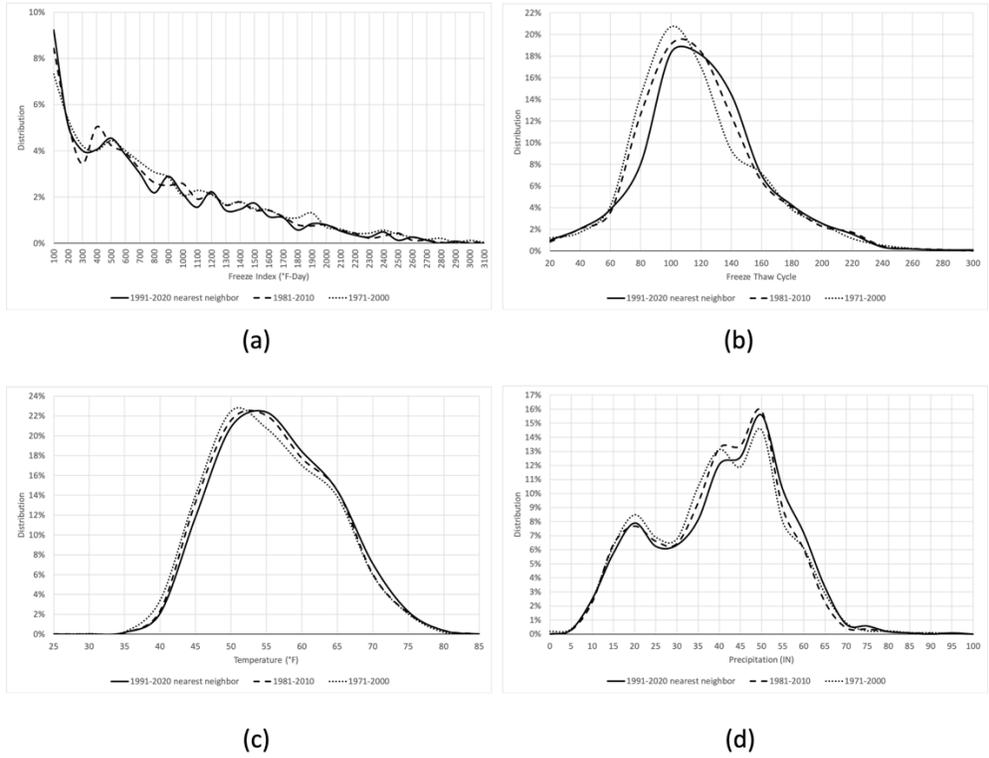

*Figure 6: (a) Freeze index trend comparison; (b) Freeze-thaw cycle trend comparison; (c) Temperature trend comparison; (d) Precipitation trend comparison*

Figure 8 presents the climate attributes in the NOAA regions. Table 1 provides the ANOVA result for the NOAA regions and identifies climate attributes that have statistically significant differences.

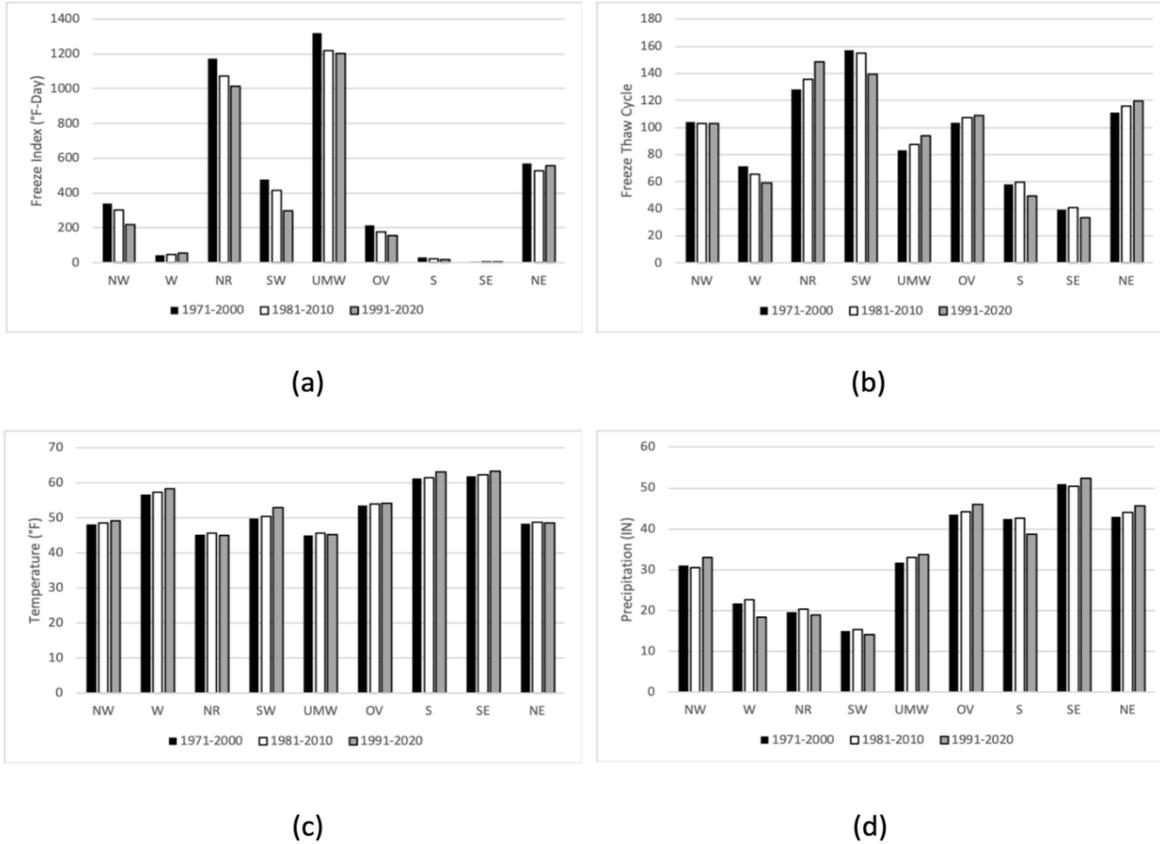

*Figure 7: Average annual freeze index, NOAA regions; (b) Average freeze-thaw cycle, NOAA regions; (c) Average temperature, NOAA regions; (d) Average precipitation, NOAA regions*

*Table 1: ANOVA Results for NOAA Regions: P-Value Tabulation*

|  | Climate Attributes | | | |
| --- | --- | --- | --- | --- |
| NOAA climate regions | Freeze Index | Freeze-Thaw Cycle | Precipitation | Temperature |
| Northwest | 0.5530 | 0.7640 | 0.9480 | 0.6496 |
| West | 0.6750 | 0.7930 | 0.6040 | 0.6450 |
| Northern Rockies and Plains | 0.6987 | 0.5727 | 0.2097 | 0.7949 |
| Southwest | 0.5695 | 0.6645 | 0.2640 | 0.5592 |
| Upper Midwest | 0.3130 | **<0.0001** | **0.0011** | 0.6990 |
| Ohio Valley | **0.0454** | 0.3904 | **0.0001** | 0.0689 |
| South | 0.1576 | 0.1823 | 0.4993 | 0.1703 |
| Southeast | **0.0283** | **0.0229** | **0.0011** | 0.2019 |
| Northeast | 0.4192 | 0.4143 | **0.0367** | 0.3252 |

Note: bold numbers indicate statistical significant difference between the means of climate attribute across the two time windows

Comparing Figure 5(a) with previous study shows that the freeze index line has been moving steady away from the west coast and the southwestern states. California, Nevada, Arizona and New Mexico see the freeze line moving further away from the south, while Oregon and California see the freeze index line moving away from the coastal region. The maximum freeze indices still occur in northern Minnesota, northern Maine, and northeastern North Dakota, where the freeze index is above 2,500°F-days. The Rocky Mountain region remains an area of high freeze index, with values as high as 2,000°F-days, specifically in western Wyoming and the central Colorado region.

From figure 5(b), it can be seen that Rocky Mountain regions remain areas with high freeze-thaw cycles (>200), with the freeze-thaw cycle starting to decrease near the Pacific coastal areas in Washington, Oregon, and California. The midwestern states remain areas of high freeze-thaw cycles, with values between 118-150 in parts of Missouri, Illinois, Indiana, Ohio, West Virginia, and Pennsylvania. In general, freeze-thaw cycle trends did not change much from 1981-2010 and 1991-2020.

Figure 6 depicts the climate attributes and their respective trends in 1991-2020. Because the 1991-2020 climate normal data used in this study contains three times as many data as the previous study, more details were uncovered regarding the spatial climate patterns. Figure 6(b) shows the freeze-thaw cycle peak around 100-120, however, there is an additional smaller peak around 200, a detail that was not shown in the previous study. Figure 6(d) shows the precipitation having two peaks, one around 15-20 inches and a second peak around 50 inches, this was not shown in the previous study either.

Figure 7 depicts comparisons across three time windows using only the weather stations that are present across the three time windows. In cases where the weather stations are no longer present, the geographically nearest weather station was used as a substitute. For freeze index, there are more variabilities in 1991-2020 compare to the previous two time windows, with peaks at approximately 500°F-days, 900°F-days,1,200°F-days, and 1,500°F-days; there is also an increase in freeze index around 100°F-days in the past three decades, from approximately 7% in 1971-2000 to approximately 9% in 1991-2020. For freeze-thaw cycle, there is an increase in higher freeze-thaw cycle frequency around 120-160 cycles. Average temperature has shown an increase from around 50°F in 1971-2000 to around 55°F in 1991-2020. Precipitation remains relatively constant during these time windows.

Not all regions experience the same degree of changes. There was a decrease in freeze index in all NOAA regions, except for the West, where freeze index has been increasing, and the Northeast, where freeze index has remained relatively constant. Freeze-thaw cycle has fluctuated across all regions. Freeze-thaw cycles increased in Northern Rockies, Upper Midwest, Ohio Valley, and Northeast regions, but it is on a declining trend in the West, Southwest, South, and Southeast regions; freeze-thaw cycle remained relatively constant in the Northwest region. Temperature increased or remained relatively constant across all nine regions. Finally, precipitation increased in the Northwest, Upper Midwest, Ohio Valley, Southeast, and Northeast regions, while it decreased in the West, Southwest, and South regions, and at the same time remaining relatively constant in the Northern Rockies region.

Most changes that are observed were not statistically significant at a 5% level of significance. Table 1 indicates that only the Upper Midwest, Ohio Valley, Southeast, and Northeast regions had statistically significant changes in some of the climate attributes. The Upper Midwest experienced statistically significant changes in freeze-thaw cycle and precipitation; Ohio Valley experienced statistically significant changes in freeze index and precipitation, Southeast experienced statistically significant changes in freeze index, freeze-thaw cycle, and precipitation, and Northeast experienced statistically significant changes in precipitation only.

## DISCUSSION AND CONDLUDING REMARKS

The updated freeze index and freeze-thaw cycle isarithmic maps represent the most comprehensive and up to date climate attribute spatial trends for the continental United States. With three times as many weather stations incorporated in this follow up study, we were able to observe more details and have a better understanding of the trends than the previous study. The previous study used roughly one weather station per county, while this was adequate for the eastern and midwestern states, it posed a slight problem for the western states, where the counties are geographically larger. The current study was able to use more than one weather station per county, thus significantly improving the resolution of the isarithmic maps. Therefore, when using the current data to produce isarithmic maps on the state level, there would be far more details present than the previous dataset because of the increased spatial resolution. The maps presented in this study, along with any potential maps that would be produced on the state level, would be ideal for incorporating into the mechanistic-empirical pavement design process, and for determining best-practice strategies for design, construction, operation, and maintenance of civil roadway infrastructures.

Most changes in the climate attributes were found to be statistically insignificant for most of the nine NOAA regions. Four of the nine regions had statistically significant changes. Both Upper Midwest and Ohio Valley regions experienced statistically significant changes in two of the four climate attributes (freeze-thaw cycle and precipitation in Upper Midwest, and freeze index and precipitation in Ohio Valley); in the previous study, Upper Midwest experienced statistically significant changes in all four climate attributes, while Ohio Valley had statistically significant changes in all climate attributes except for precipitation. Southeast experienced statistically significant changes in all climate attributes except for temperature, Southeast did not have statistically significant changes in any climate attribute in the previous study. Northeast experienced statistically significant change in precipitation, which differs from the previous study, where Northeast only experienced statistically significant change in freeze-thaw cycle. As shown in the previous study and this follow-up study, continual updating of climate information is a necessary and useful in order to capture any changes in the climate attributes.

The updated maps provide a better picture of the freezing season characteristics in states that were previously shown not to be affected by winter weather conditions, and indicates the change in boundaries for both freeze index and freeze-thaw cycle occurrences. These new maps will not only give practicing engineers more current values to consider during the design process, they will also be important when developing future infrastructure maintenance procedures during the winter season by providing accurate assessment of the regional climate conditions.